# Fermi level dependent charge-to-spin current conversion by Dirac surface state of topological insulators

K. Kondou[1*], R. Yoshimi[2], A. Tsukazaki[3], Y. Fukuma[1,4], J. Matsuno[1], K. S. Takahashi[1], M. Kawasaki[1,2], Y. Tokura[1,2] and Y. Otani[1,5]

[1] *RIKEN Center for Emergent Matter Science (CEMS), Wako 351-0198, Japan*

[2] *Department of Applied Physics and Quantum-Phase Electronics Center (QPEC), University of Tokyo, Tokyo 113-8656, Japan*

[3] *Institute for Materials Research, Tohoku University, Sendai 980-8577, Japan*

[4] *Frontier Research Academy for Young Researchers, Kyushu Institute of Technology, Iizuka 820-8502, Japan*

[5] *Institute for Solid State Physics, University of Tokyo, Kashiwa 277-8581, Japan,*

**\*Correspondence to: kkondou@riken.jp**

The spin-momentum locking at the Dirac surface state of a topological insulator (TI)[1-6] offers a distinct possibility of a highly efficient charge-to-spin current (C-S) conversion compared with spin Hall effects in conventional paramagnetic metals[7-13]. For the development of TI-based spin current devices, it is essential to evaluate its conversion efficiency quantitatively as a function of the Fermi level $E_F$ position. Here we exemplify a coefficient of $q_{ICS}$ to characterize the interface C-S conversion effect by using spin torque ferromagnetic resonance (ST-FMR) for $(Bi_{1-x}Sb_x)_2Te_3$ thin films whose $E_F$ is tuned across the band gap. In bulk insulating conditions, interface C-S conversion effect via Dirac surface state is evaluated as nearly constant large values of $q_{ICS}$, reflecting that the $q_{ICS}$ is inversely proportional to the Fermi velocity $v_F$ that is almost constant. However, when $E_F$ traverses through the Dirac point, the $q_{ICS}$ is remarkably suppressed possibly due to the degeneracy of surface spins or instability of helical spin structure. These results demonstrate that the fine tuning of the $E_F$ in TI based heterostructures is critical to maximizing the efficiency using the spin-momentum locking mechanism.

Three dimensional topological insulators (TIs) possess metallic surface states in which spins of carriers are locked orthogonal to their momenta owing to time-reversal invariant. This feature is called "spin-momentum locking" which has been employed as a principal mechanism to induce spin accumulation in the surface states of TIs[1,3-5,14-17]. Conceptually, charge current can fully contribute to spin current via the spin-momentum locking; 100% C-S conversion efficiency $\theta_{CS}$ is expected at the non-TI / TI heterointerface. This efficient C-S conversion can be widely applicable to spintronic devices. However, in reality, the C-S conversion efficiency deduced by the spin torque measurement has been above 100% for TIs with mixed contribution from the surface and bulk bands[3,4], when the efficiency is defined as $\theta_{CS} = J_S / J_C$, where $J_S$ is spin current density (Am$^{-2}$) and $J_C$ is charge current density (Am$^{-2}$) in whole TI layer. Here we distil the contribution from the Dirac electrons in the C-S conversion process and clarify the role of Fermi level $E_F$ and Fermi velocity $v_F$ employing TI samples with various $E_F$ positions. Accordingly, we define the interfacial C-S conversion coefficient $q_{ICS}$ as $q_{ICS} = J_S / j_C$, where $j_C$ is surface charge current density (Am$^{-1}$). In this study, we quantitatively evaluate the interface C-S conversion effect by ST-FMR technique for (Bi$_{1-x}$Sb$_x$)$_2$Te$_3$ / Cu / Ni$_{80}$Fe$_{20}$ (Py) tri-layer films as shown in Fig. 1a. With insertion of a Cu layer between TI and ferromagnet layers, spin accumulation at the surface states can be exclusively evaluated owing to suppression of the exchange coupling between ferromagnet and the surface states of TI[18-20].

The ST-FMR technique has been usually employed to evaluate an induced spin current via spin Hall effect in paramagnetic metals[3,10]. Here we apply this technique to characterize quantitatively interface C-S conversion effect via spin-momentum locking at the Cu-inserted TI-based tri-layer heterostructures as shown in the top schematic of Fig. 1a. A photo of the device and measurement circuit is drawn in the bottom of Fig. 1a. To evaluate $q_{ICS} = J_S / j_C$ by ST-FMR, charge current distribution in the tri-layer should be numerically clarified, because $j_C$ in TI layer is one of the dominant factors for this evaluation technique. When an rf current flows in the tri-layer film, FMR is excited in the top Py layer under the external static magnetic field $H_{ext}$. Owing to the presence of highly conductive Cu layer, the peak of the current density is located towards the outside of the Py layer so that homogeneous rf fields ($H_{rf}$) can be applied to the Py layer (see Supplementary information S1), providing a better characterization condition for C-S conversion effect by ST-FMR[21]. The spin accumulation simultaneously takes place in the surface state of TI, the accumulated spins generate a spin current $J_S$ at orthogonal direction diffusing into both Cu and Py layers and then exert spin torque on the Py layer (white arrow in Fig. 1a). A typical ST-FMR spectrum is shown in Fig. 1b: the symmetric voltage $V^{Sym}$ is attributable to the spin torque $\tau_{//}$ originating from spin current density $J_S$ (details discussed later). By the quantitative evaluation of the $V^{Sym}$, we can deduce the interface C-S conversion coefficient $q_{ICS}$.

From the concept of spin-momentum locking, the magnitude of $J_S$ is governed by that of $j_C$ linked with the conductivity of the surface states on TI layer depending on Fermi energy and Dirac dispersion: the Fermi velocity $v_F$ and the Fermi wave vector $k_F$ (refs. 15-17). Therefore in this study, we prepared $(Bi_{1-x}Sb_x)_2Te_3$ (BST) thin films with systematically tuned Fermi levels by $x$ to investigate the relationship between the $q_{ICS}$ and the transport properties at the surface state. Figure 2a shows the Hall coefficients $R_H$ obtained for 7 films having different Sb composition $x$ at 10 K. The value of $R_H$ is negative for $x = 0$ and its magnitude increases while the $x$ is increased to 0.82, indicating $n$-type conduction and reduction of electron density. The polarity of $R_H$ abruptly reverses its sign when the $x$ reaches the value around 0.84, revealing that the Fermi energy traverses Dirac point (DP). In the range $0.88 \leq x \leq 1$, the polarity of $R_H$ is positive and hence $p$-type conduction is evident. The mobility $\mu$ in BST films with $x = 0.88$ reaches a maximum value of 1900 cm$^2$/Vs, which is comparable to the previous results[22]. These transport properties assure that the Fermi level of BST is systematically varied from $n$- to $p$-type across the DP in a controlled manner, as shown in Fig. 2d.

Figure 2c shows the Sb composition $x$ dependence of $V^{Sym}$ of ST-FMR spectrum. The sign of $V^{Sym}$ indicates the spin polarization direction of the spin current. We found the positive $V^{Sym}$ in both $n$- and $p$-type BST films, which is an ideal feature of the C-S conversion via spin-momentum locking[1-6]. In Dirac dispersion drawn in Fig. 2e and 2f, spins on the Fermi circles of $n$- and $p$-type surface states of BSTs are whirling clockwise and anti-clockwise, respectively[24]. When electric field $E_x$ is applied towards the $-x$ direction, the Fermi circle with the chiral spin structure is shifted from the dashed to solid line circles in proportion to $E_x$ along $k_x$ as shown in Fig. 2e. When the Fermi level

5 / 20

$E_F$ is above DP, the surface state of BST films is more populated by down spins, generating the spin polarization of the spin current along $-y$ direction. When the $E_F$ is in the valence band of Dirac dispersion, up spins with momenta along $+k_x$ are less than down spins with $-k_x$. Thus the accumulated spins are oriented along the same direction for both *n*- and *p*-type BST films. Note that these results are different from the case of a typical semiconductor such as GaAs[23], whose spin Hall effect exhibits a different sign depending on the carrier type.

Figure 3 shows the *x* dependence of $q_{ICS}$ and the spin current conductivity $\sigma_S$ of BST films. The value of $q_{ICS}$ can be experimentally evaluated from the ratio of $V^{Sym}$ to $V^{Anti}$ in ST-FMR spectrum. In the ST-FMR process, the $V^{Sym}$ and $V^{Anti}$ correspond to the spin torques $\tau_{//}$ and the field torque $\tau_{\perp}$ due to an Oersted field generated by charge current flow. These two torques per unit moment on the Py are respectively expressed as $\tau_{//} = \hbar J_S/(2e\mu_0 M_S t_{Py})$ and $\tau_{\perp} = \xi \{J_C^{Cu} t_{Cu}/2 + j_C/2\}$, where $M_S$, $t$, and $\xi$ are saturation magnetization, film thickness, and reduction factor of rf field. The value of $\xi$ is calculated numerically by means of finite element method (See Supplemental information S1). The $q_{ICS}$ can thus be given by

$$q_{ICS} \equiv \left(\frac{J_S}{j_C}\right) = \left(\frac{\tau_{//}}{\tau_{\perp}}\right) \frac{a\xi t_{Py} t_{Cu} e\mu_0 M_S \{1 + M_S/H_{ext}\}^{0.5}}{\hbar t_{BST}}$$

$$= \left(\frac{V^{Sym}}{V^{Anti}}\right) \frac{a\xi t_{Py} t_{Cu} e\mu_0 M_S \{1 + M_S/H_{ext}\}^{0.5}}{\hbar t_{BST}} \quad (1),$$

where *a* is the ratio of $J_C^{Cu}$ (Am$^{-2}$) to $j_C$. The spin current density into Py $J_S^{Py}$ (Am$^{-2}$) is proportional to spin accumulation at surface state of TI, $\langle \delta S_0 \rangle$, which is expressed as

6 / 20

$$\langle \delta S_0 \rangle = \frac{\hbar}{2} k_F \delta k_x = \frac{e k_F E_x \tau}{2} = \frac{\mu k_F^2 \hbar E_x}{2 v_F} \quad (2),$$

where $k_F$ is the Fermi wave number, $\delta k_x$ the shift of Fermi circle, $\tau$ the relaxation time. In the 2D system, $k_F^2$ is proportional to the carrier density. Therefore, $\langle \delta S_0 \rangle$ is reduced to $\hbar j_C / 2e v_F$ and $q_{ICS} = J_S^{Py} / j_C \propto v_F^{-1}$ is obtained. According to angle-resolved photoemission spectra, $v_F$ in BST slightly increased from 3.6 to 3.9 × $10^5$ m/s with the Sb composition $x$ from 0.5 to 0.9 (ref. 16), implying that $q_{ICS}$ is almost constant. For bulk insulating BST films with $x$ = 0.5, 0.7 and 0.9, we observed that the values of $q_{ICS}$ are lying within a range of 0.45 to 0.57 nm$^{-1}$, similar in magnitude to the previous results of ST-FMR measurements in Py/Bi$_2$Se$_3$ bilayer film[3]. Since we assume the surface thickness to be 1-nm, the $\theta_{CS}$ is estimated to be 45 to 57% for these BST films, yielding much higher conversion efficiency than those in typical transition metals like $\beta$-Ta (15%) and $\beta$-W (33%)[10-13]. Consequently, the interface C-S conversion effect via spin-momentum locking at the surface state is fairly evaluated with $q_{ICS}$, manifesting consistency to the naive expectation of 100% conversion efficiency. Note that the estimated conversion efficiency with $\theta_{CS}$ is proportional to the conducting channel thickness and therefore large values of $\theta_{CS}$ claimed for high conversion efficiencies in previous studies[3,4] may be overestimated with assuming thicker conducting layer. Conversely, the expectation of 100% C-S conversion means that the surface states conduction may be contributed from 2-nm region.

In contrast to the almost constant $q_{ICS}$ for the bulk insulating BST films with $x = 0.5$, 0.7 and 0.9, the values of $q_{ICS}$ show a sharp dip around DP $x \sim 0.84$; the $q_{ICS}$ dramatically decreases when $E_F$ locates around DP or equivalently $k_F$ becomes about zero, where the spin-momentum locking disappears (see Fig. 2d). In such a band-crossing point, a finite scattering rate of the Dirac electrons may mix the spin-polarized dispersions to cause some imbalance between up and down spin populations, yet it is likely that a marked reduction of $\langle \delta S_0 \rangle$ takes place as observed. For bulk conductive BST films with $x = 0$ ($Bi_2Te_3$) and 1 ($Sb_2Te_3$), values of $q_{ICS}$, estimated with the assumption of 1 -nm surface layer, are found to be roughly equal to or even twice as much as that for the bulk insulating BST films. However, quantitative evaluation of $q_{ICS}$ and $\theta_{CS}$ in bulk conducting samples is quite difficult due to the following reasons. The estimation of parasitic current effect: here we assume that the surface is as conductive as the bulk (See Supplemental information S1). Considering that the surface is expected to be more conductive, the present value of $q_{ICS}$ might be overestimated. Rashba effect: if the charge current in bulk band also contributes to the $J_S$ via Rashba split band like $Bi_2Se_3$ (ref. 17), the opposite spin polarization may cancel a part of surface spin accumulation (See Supplemental information S3). Bulk spin Hall effect: the bulk charge current can also contribute to the $J_S$ via ordinary spin Hall effect, of which sign is yet unclear so far. Finally we show the spin current conductivity $\sigma_S$, defined as $\sigma_S = q_{ICS}\sigma^{surf}$, where $\sigma^{surf}$ is conductivity of surface state of TI, as a function the Sb composition $x$ in Fig. 3b. The $\sigma_S$ in bulk insulating BST excluding the DP and bulk conductive BST take close values in the range from 0.7 to $1.8 \times 10^5$ $\Omega^{-1}$m$^{-1}$, which are comparable to those reported for three dimensional processes originating from spin Hall effect in paramagnetic metals such as Pt ($3.4 \times 10^5$ $\Omega^{-1}$m$^{-1}$)[10] and $\beta$-W ($1.3 \times 10^5$ $\Omega^{-1}$m$^{-1}$)[11]. This high value of $\sigma_S$ is certainly beneficial not only for realizing highly efficient magnetization switching but also for realizing non-volatile spin switching for Boolean and non-Boolean logics initially based on metal spin Hall effects[25].

In summary, we have investigated the charge-to-spin conversion phenomena at the Dirac surface state of BST films across DP by using ST-FMR technique. For fair comparison of C-S conversion effect, $q_{ICS}$ is a better figure of merit than $\theta_{CS}$ since there is no need to account for conducting thickness effect. We found that the C-S conversion effect is strongly dependent on the $E_F$ position; when it is in bulk gap, the same sign of spin accumulation direction and the comparable amplitude of the accumulation are achieved. The C-S conversion diminishes when $E_F$ is close to DP, probably due to other effects of disorder and/or spin scattering. These findings have important implications for development of future spintronic devise using the interface spin conversion effect.

Methods

Sample fabrication

We grew 8-nm thick BST films on semi-insulating InP (111) substrates by molecular beam epitaxy. Detailed growth conditions are described in previous paper[22]. Bi/Sb ratio was tuned by the ratio of beam equivalent pressures of Bi and Sb. Resistivity and Hall effect measurements were carried out with small chips divided from the same samples used for ST-FMR measurements (see Supplemental information S2). Thin films of 8-nm Cu/10-nm $Ni_{80}Fe_{20}$ (Py) / 5-nm $Al_2O_3$ were grown on the BST films by e-beam evaporation at $5 \times 10^{-5}$ Pa. $Al_2O_3$ is used as an insulating capping layer. The resistivity of Cu and Py are measured to be 10 and 60 μΩcm at 10 K. The BST/Cu/Py tri-layer films were patterned into rectangular elements ($10 \times 30$, $15 \times 45$, $20 \times 60$, $30 \times 90$, $40 \times 120$ μm$^2$) using optical lithography and Ar-ion etching technique. A co-planar waveguide of 5-nm Ti /200-nm Au was deposited on both sides of the rectangular elements.

ST-FMR measurement set up

An rf current with an input power of 10 dBm is applied along the long edge of the rectangle by a microwave analog signal generator (Keysight: MXG N5183A). An external static magnetic field $H_{ext}$ in the range from 0 to 2.0 kOe is also applied in the film plane with an angle at $\theta = 45^o$ from current flow direction. All the experiments are performed at 10 K to measure the surface dominant properties of TI.

**Acknowledgements**

This work was supported by Grant-in-Aid for Scientific Research on Innovative Area, "Nano Spin Conversion Science" (Grant No. 26103002). R. Y. is supported by the Japan Society for the Promotion of Science (JSPS) through a research fellowship for young scientists. This research was supported by the Japan Society for the Promotion of Science through the Funding Program for World-Leading Innovative R & D on Science and Technology (FIRST Program) on "Quantum Science on Strong Correlation" initiated by the Council for Science and Technology Policy.

Author contribution

Y.O. and Y.T. conceived the project. K.K. made the devices and performed the spin torque ferromagnetic resonance measurements. R.Y grew the topological insulator thin films and performed Hall measurement. K.K. analysed the data and wrote the manuscript with contributions from all authors. A.T., Y.F., K.S.T., J.M., M.K., Y.T. and Y.O jointly discussed the results.

**Competing financial interests**

The authors declare no competing financial interests.

**Figure 1 | Spin current generation and detection in BST/Cu/Py tri-layer structure**

**a**, ST-FMR measurement circuit and device design employing BST/Cu/Py heterostructures. White arrows on the surfaces of BST layer show the polarization direction of spin accumulation. Static magnetic field ($H_{ext}$) is tilted by $\theta = 45°$. **b**, A typical ST-FMR spectrum measured for a BST ($x = 0.7$)/Cu/Py tri-layer film at 10 K. Red plots shows experimental spectrum, which can be divided into symmetric ($V^{Sym}$: green line) and anti-symmetric ($V^{Anti}$: blue line) parts. $V^{Sym}$ and $V^{Anti}$ correspond to spin current induced FMR and Oersted field induced FMR, respectively.

**Figure 2 | Transport properties of BST films and detected $V^{Sym}$ as a function of Sb composition**

**a**, Hall coefficient $R_H$ for BST films measured at 10 K. **b**, Mobility $\mu$ and carrier density $n_{2D}$, $p_{2D}$ as a function of Sb composition at 10 K. **c**, Symmetric voltage of ST-FMR as a function of Sb composition. Input rf frequency and power is 7 GHz and 10 dBm, respectively. The error bars are standard deviation in 5 samples with different dimensions. **d**, A schematic of energy dispersion in BST. **e, f** (Top) spin accumulation due to Fermi circle shift at surface state of *n*-type (**e**) and *p*-type (**f**) BST. Solid and dashed line circles are Fermi circle with $E_x$ and without $E_x$, respectively. (Bottom) Difference of Fermi distribution ($f - f_0$) by applying electric field

**Figure 3 | Sb composition dependence of charge-to-spin current conversion efficiency of BST**

**a**, Interface C-S conversion efficiency $q_{ICS}$ as a function of Sb composition. Inset shows the band structure and Fermi level position for each Sb composition. Bulk insulating BST with $0.50 \leq x \leq 0.90$ should have only surface transport. $Bi_2Te_3$ ($x = 0$) and $Bi_2Sb_3$ ($x = 1$) have both bulk and surface conduction paths. Error bars are standard deviation in 5 samples with different dimensions. **b**, spin current conductivity as a function of Sb composition.

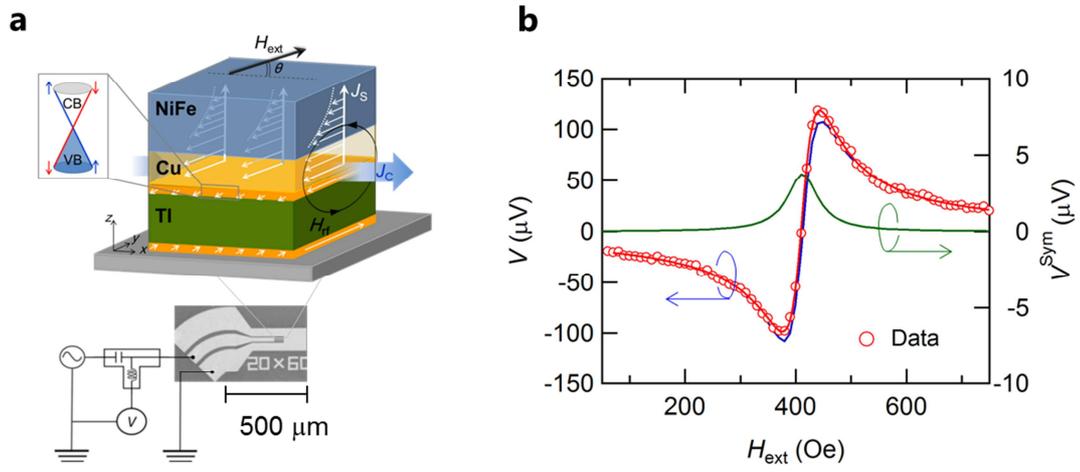

FIG. 1 K. Kondou *et al.*,

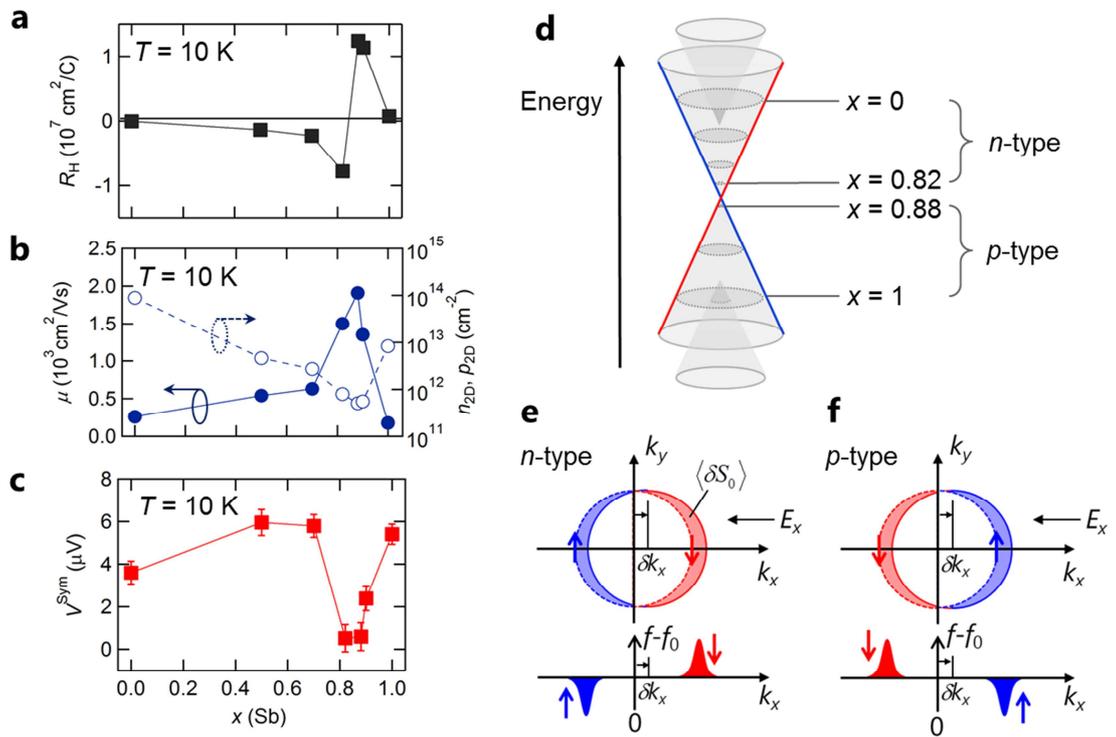

FIG. 2 K. Kondou *et al.*,

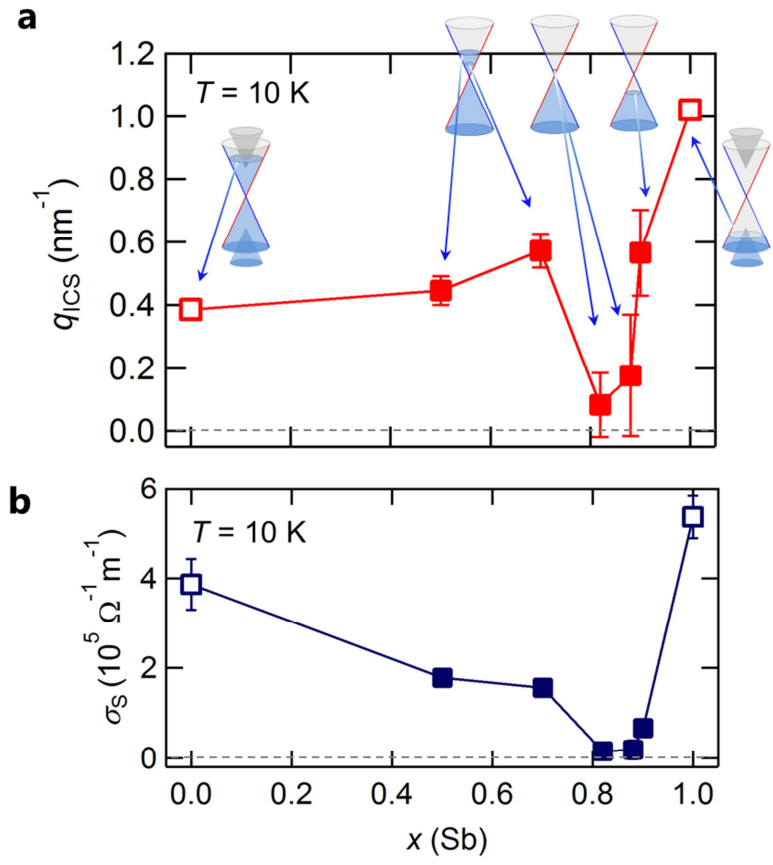

FIG. 3 K. Kondou *et al*.,